\documentclass{iopart}
\usepackage{epsfig}
\usepackage{amssymb}

\begin{document} 

\newcommand{\vk}{{\vec k}} 
\newcommand{\vK}{{\vec K}}  
\newcommand{\vb}{{\vec b}}  
\newcommand{\vp}{{\vec p}}  
\newcommand{\vq}{{\vec q}}  
\newcommand{\vQ}{{\vec Q}} 
\newcommand{\vx}{{\vec x}} 
\newcommand{\vh}{{\hat{v}}} 
\newcommand{\cO}{{\cal O}}
\newcommand{\be}{\begin{equation}} 
\newcommand{\ee}{\end{equation}}  
\newcommand{\half}{{\textstyle\frac{1}{2}}}  
\newcommand{\gton}{\mathrel{\lower.9ex \hbox{$\stackrel{\displaystyle 
>}{\sim}$}}}  
\newcommand{\lton}{\mathrel{\lower.9ex \hbox{$\stackrel{\displaystyle 
<}{\sim}$}}}  
\newcommand{\ben}{\begin{enumerate}}  
\newcommand{\een}{\end{enumerate}} 
\newcommand{\bit}{\begin{itemize}}  
\newcommand{\eit}{\end{itemize}} 
\newcommand{\bc}{\begin{center}}  
\newcommand{\ec}{\end{center}} 
\newcommand{\bea}{\begin{eqnarray}}  
\newcommand{\eea}{\end{eqnarray}}

\title{Charm elliptic flow from quark coalescence dynamics}
 
\date{\today}
 
\author{D\'enes Moln\'ar}
\address{Department of Physics, Ohio State University, Columbus, OH 43210}


\begin{abstract}
From covariant transport theory, 
a significant $\sim 10$\% light quark elliptic flow at RHIC implies
an elliptic flow of similar magnitude for charm quarks,
at moderately large $p_T > 2.5-3$ GeV.
At lower transverse momenta, charm quark elliptic flow reduces 
progressively, reminiscent of the mass ordering pattern in ideal
hydrodynamics.
From the quark flows we predict the elliptic flow of $D$ mesons at RHIC
via quark coalescence.
The large parton opacities needed to generate the light quark flow
also lead to substantial $\sim 40-50$\% secondary charm production.
\end{abstract}



\section{Introduction}
Elliptic flow~\cite{flow-review},  $v_2 \equiv\langle \cos(2\phi)\rangle$,
the second Fourier moment of the azimuthal momentum distribution,
is an important experimental probe 
that provides information about the opacity~\cite{Binv2,v2,pQCDv2,coalv2} 
of the nuclear medium created in noncentral $A+A$ reactions.
Heavy flavor 
elliptic flow is especially interesting because 
it can tell to what degree heavy quarks interact and thermalize.
Though heavy quarks experience the same partonic environment as light partons,
their large mass is expected to hamper equilibration,
at least at RHIC energies $\sqrt{s_{NN}}\sim 100-200$ GeV.
Studying the 
breakdown of equilibrium for the most abundant {\em charm} 
quarks may shed more light on the origin
of the remarkable thermalization apparent
in the light sector~\cite{thermalFits,Kolbhydro,Teaneyhydro,Hiranohydro}.

In principle the experimentally more accessible
heavy flavor spectra can also constrain dynamical scenarios.
However, surprisingly, current (indirect) measurements~\cite{PHENIXe}
from $Au+Au$ at RHIC are compatible with 
{\em both} perturbative production without charm rescattering 
and complete charm equilibration~\cite{Batsouli}.
Charm hadron, e.g., $D$ meson, 
elliptic flow will provide a more decisive experimental test~\cite{Kaneta}.

Though 
several schematic scenarios have been explored~\cite{charmcoal,Texascharm},
up to now there is no quantitative calculation of charm elliptic flow
in heavy ion collisions at RHIC.
The aim of this study is to make predictions for 
$Au+Au$ at $\sqrt{s_{NN}}=200$ GeV
based on covariant parton transport 
theory~\cite{Binv2,v2,ZPC,nonequil,inelv2,hbt,dyncoal,XuGreiner}.
That approach successfully describes the magnitude and saturation of 
$v_2(p_T)$ in the light sector,
but requires initial parton densities and/or cross sections that are much
larger than perturbative QCD estimates~\cite{v2}.
Large opacities are also indicated by
pion interferometry data~\cite{hbt}.
Nevertheless, the origin of such opaque conditions is a puzzle.

One scenario that alleviates the opacity problem is 
hadronization via quark 
coalescence~\cite{ALCORMICOR,VoloshinQM02,Texascoal,Dukecoal,Hwacoal,coalv2,%
dyncoal}.
In the coalescence process mesons form from a constituent quark and antiquark,
while (anti)baryons from three (anti)quarks.
Because comoving constituents are strongly favored, elliptic flow can be 
amplified~\cite{VoloshinQM02,coalv2,charmcoal,dyncoal,coalscaling,Kolbcoal},
reducing the opacities needed~\cite{coalv2} to explain the data.
Utilizing the formulas in~\cite{charmcoal}, 
we present predictions for the elliptic flow 
of $D$ and $D_s$ mesons at RHIC.

\section{Covariant parton transport theory}

We consider here an inelastic {\em extension} 
of the Lorentz-covariant parton transport theory in
Refs.~\cite{Binv2,v2,ZPC,nonequil,hbt},
in which the on-shell parton phase space densities
$\{ f_i(x,\vp)\}$
evolve with elastic $2\to 2$ and {\em inelastic} $2\to 2$ rates as
\bea
\fl
p_1^\mu \partial_\mu f_{1,i} &=& 
\frac{1}{16\pi^2}\sum\limits_{jk\ell} 
\int\limits_2\!\!\!\!
\int\limits_3\!\!\!\!
\int\limits_4\!\!
\left(
f_{3,k} f_{4,\ell} \frac{g_i g_j}{g_k g_\ell} - f_{1,i} f_{2,j}
\right)
\left|\bar{\cal M}_{12\to 34}^{ij\to k\ell}\right|^2 
\delta^4(p_1+p_2-p_3-p_4)
\nonumber \\
\fl
&& + S_i(x, \vp_1) \ .
\label{Eq:Boltzmann_22}
\eea
$|\bar{\cal M}|^2$ is the polarization averaged scattering matrix 
element squared,
the integrals are shorthands
for $\int\limits_a \equiv \int d^3 p_a / (2 E_a)$,
$g_i$ is the number of internal degrees of freedom for species $i$,
while $f_{a,i} \equiv f_i(x, \vp_a)$.
The source functions $\{S_i(x,\vp)\}$ specify the initial conditions.

Though, in principle, (\ref{Eq:Boltzmann_22}) could be generalized for bosons
and fermions,
or inelastic $3\leftrightarrow 2$ processes~\cite{inelv2,XuGreiner},
no practical algorithm yet exists (for opacities at RHIC)
to handle the new nonlinearities such extensions introduce.
We therefore limit our study to quadratic dependence of the collision
integral on $f$.

We apply (\ref{Eq:Boltzmann_22}) to a system of massless gluons ($g=16$), 
massless light ($u$,$d$) and strange quarks/antiquarks, 
and charm quarks/antiquarks ($g=6$) with mass $M_c = 1.2$ GeV.
All  elastic and inelastic $2 \to 2$ QCD processes were taken into account: 
$gg\to gg$, 
$gq \to gq$, $qq \to qq$, $\bar q\bar q \to \bar q\bar q$,
$q \bar q \to q \bar q$; $q\bar q \to gg$, $gg\to q \bar q$, and
$q_i \bar q_i \to q_j \bar q_j$.
The matrix elements for massive quarks were taken from~\cite{Combridge}.
As in Refs.~\cite{Binv2,v2},
only the most divergent parts of the matrix elements 
were considered, regulated using a Debye mass of $\mu_D = 0.7$ GeV.

Thus, for all elastic scatterings,
$d\sigma/dt \sim 1/t^2 \to 1/(t-\mu_D)^2$,
and for $gg\leftrightarrow q\bar q$ with massless quarks,  
$d\sigma/dt \sim 1/(ut) \to 
1/(u-\mu_D^2)(t-\mu_D^2)$.
For charm, $q\bar q \to c \bar c$  to 
first approximation can be considered isotropic, 
while $gg \leftrightarrow c\bar c$ can be well approximated with
the angular dependence $d\sigma/dt \sim 1/[1-(1-4M_c^2/s)\cos^2\theta_{cm}]^2$.

The total cross section for $gg\to gg$ was taken to be constant, 
neglecting its weak logarithmic energy 
dependence. 
Its value fixes the total cross sections for all other elastic 
channels: 
\be
\sigma_{gq\to gq} = (4/9) \sigma_{gg\to gg} \qquad , \qquad
\sigma_{qq\to qq} = (4/9)^2 \sigma_{gg\to gg} \ .
\label{el_xsec}
\ee
It also determines the inelastic total cross sections, 
which on the other hand, decrease at high energy. 
For processes with only massless partons, 
the energy dependence is through the ratio 
$ r \equiv \mu_D^2/s$,
while for charm, $r$ and $R\equiv M_c^2 / s$ both play a 
role~\cite{Combridge}:
\bea
\fl
\sigma_{gg\to q\bar q} &=&
\frac{2\, r}{27} \frac{1+r}{1+2 r}\ln(1+\frac{1}{r})\, \sigma_{gg\to gg} 
\qquad , \qquad
\sigma_{q_i \bar q_i \to q_j\bar q_j} = \frac{16\, r}{243} \,
\sigma_{gg\to gg} \nonumber \\
\fl
\sigma_{gg \to c\bar c} &=& \frac{2\, r}{27}
\Theta(1-4 R) \left[
(1+4R + R^2) \ln\frac{1+\sqrt{1-4R}}{1-\sqrt{1-4R}}
-(7+3 R) \frac{\sqrt{1-4R}}{4}\right]\, \sigma_{gg\to gg}
\nonumber \\
\fl
\sigma_{q\bar q \to c\bar c} &=& \frac{16\, r}{243}
\Theta(1-4 R) (1 + 2 R) \sqrt{1 - 4R} \, \, \sigma_{gg\to gg}
\label{inel_xsec}
\eea

In order to generate sufficient~\cite{v2,coalv2} 
elliptic flow at RHIC, we take $\sigma_{gg\to gg} = 10$ mb, 
about three times the perturbative QCD estimate.
As discussed in \cite{coalv2}, this choice approximates
contributions from $2\leftrightarrow 3$ 
inelastic processes~\cite{inelv2}, and
also relies on the 
amplification of elliptic flow during hadronization via
the coalescence process.

The parton initial conditions for
$Au+Au$ at $\sqrt{s}=200A$ GeV at RHIC
with $b=8$ fm ($\approx 30$\% central) were the same as in \cite{dyncoal},
except that initial charm production was, of course, included.
Leading order pQCD minijet three-momentum distributions 
were used (with a $K$-factor of 2, GRV98LO PDFs, and $Q^2{=}p_T^2$, while
$Q^2 = \hat s$ for charm).
The low-$p_T$ divergence in the light-parton jet cross sections 
was regulated via a smooth extrapolation
below $p_\perp < 2$ GeV 
to yield a total parton $dN(b{=}0)/dy=2000$ at midrapidity.
This choice is motivated by the observed $dN_{ch}/dy \sim 600$ and the 
expectation that hadronization is dominated by quark 
coalescence~\cite{VoloshinQM02,Texascoal,Dukecoal,coalv2,dyncoal}.
The transverse density distribution was proportional
to the binary collision
distribution for two Woods-Saxon distributions,
therefore $dN^{parton}(b{=}8\ {\rm fm})/dy \approx 500$.
Perfect $\eta=y$ correlation was assumed.

The transport solutions were obtained via Molnar's Parton Cascade 
algorithm~\cite{MPC} (MPC),
which employs the parton subdivision technique~\cite{subdivision} 
to maintain Lorentz covariance and causality.

\section{Results for charm quarks}

\begin{figure}[hbpt] 
\begin{center}
\hspace*{-0.2cm}%
\epsfig{file=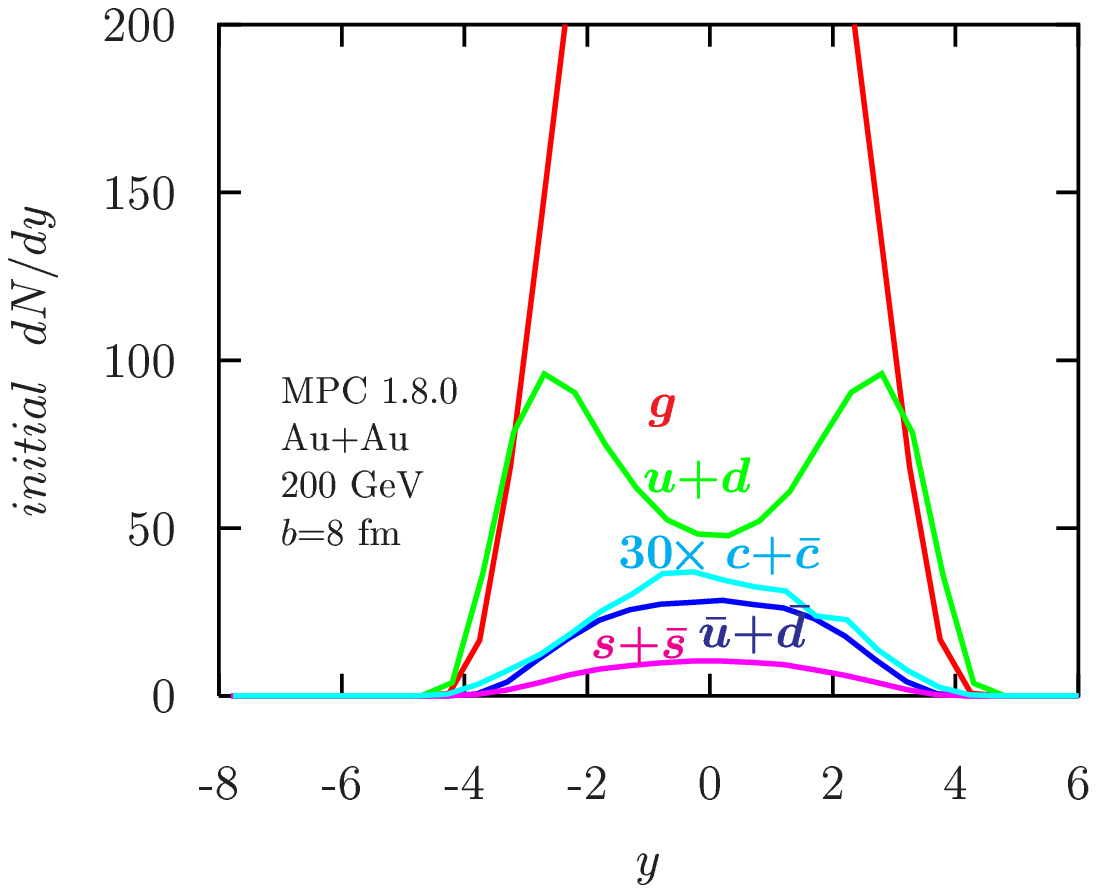,height=2.1in,width=2.6in,clip=5,angle=0} 
\hspace*{-0.2cm}%
\epsfig{file=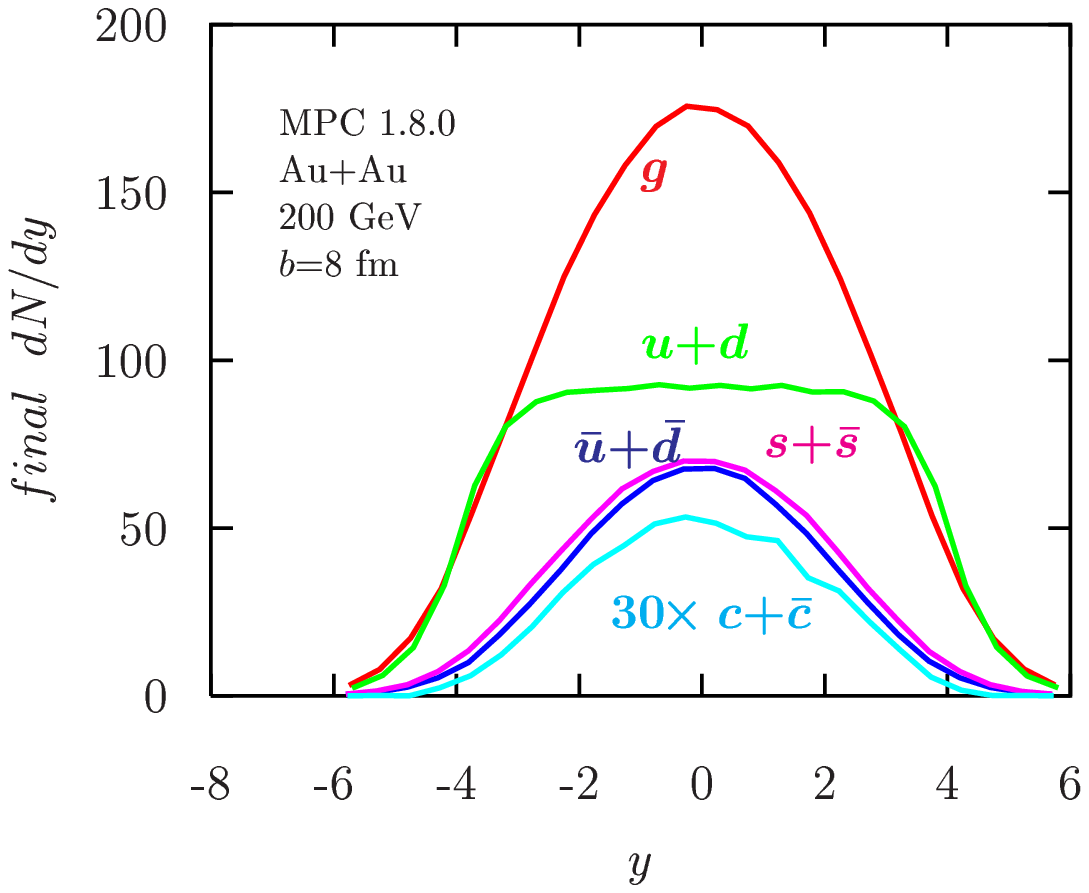,height=2.1in,width=2.6in,clip=5,angle=0} 
\end{center}
\vspace*{-0.2cm} 
\caption{\label{fig1}
Initial (left panel) and final (right panel) parton rapidity distributions
in $Au+Au$ at $\sqrt{s_{NN}}=200$ GeV with $b=8$ fm at RHIC, 
computed from the transport model
MPC~\cite{MPC}. 
}
\end{figure}

Because of the inelastic channels, (\ref{Eq:Boltzmann_22})
is suitable for studying flavor equilibration at RHIC
(with the limitation that the total parton number is fixed).
Results for the initial and final rapidity distributions
of each parton species are shown in
the left and right panels of Fig.~\ref{fig1}, respectively.
The initial condition is dominated by gluons, 
roughly half of which fuse to $q-\bar q$ pairs.
This fills in the dip in the light ($u$, $d$) 
quark distributions near midrapidity,
doubles the light antiquark distributions ($\bar u$, $\bar d$),
enhances strangeness about five-fold, and increases charm by $40-50$\%.
Flavor changing processes involving only quarks
$q_i \bar q_i \to q_j \bar q_j$ influence the flavor chemistry by 
only $10-15$\%.

These results show a high degree of light favor and strangeness
equilibration at RHIC.
The final light antiquark ($\bar u$, $\bar d$) 
and strange quark distributions ($s$, $\bar s$) are 
nearly equal, though are still below the light quark 
($u$, $d$) yields.
The baryon number distribution, which is
one-third of the $u+d - (\bar u + \bar d)$ difference,
does not vanish even at midrapidity, showing that a baryon free 
region is not realized at RHIC energies.

\begin{figure}[hbpt] 
\center
\hspace*{-0.2cm}%
\epsfig{file=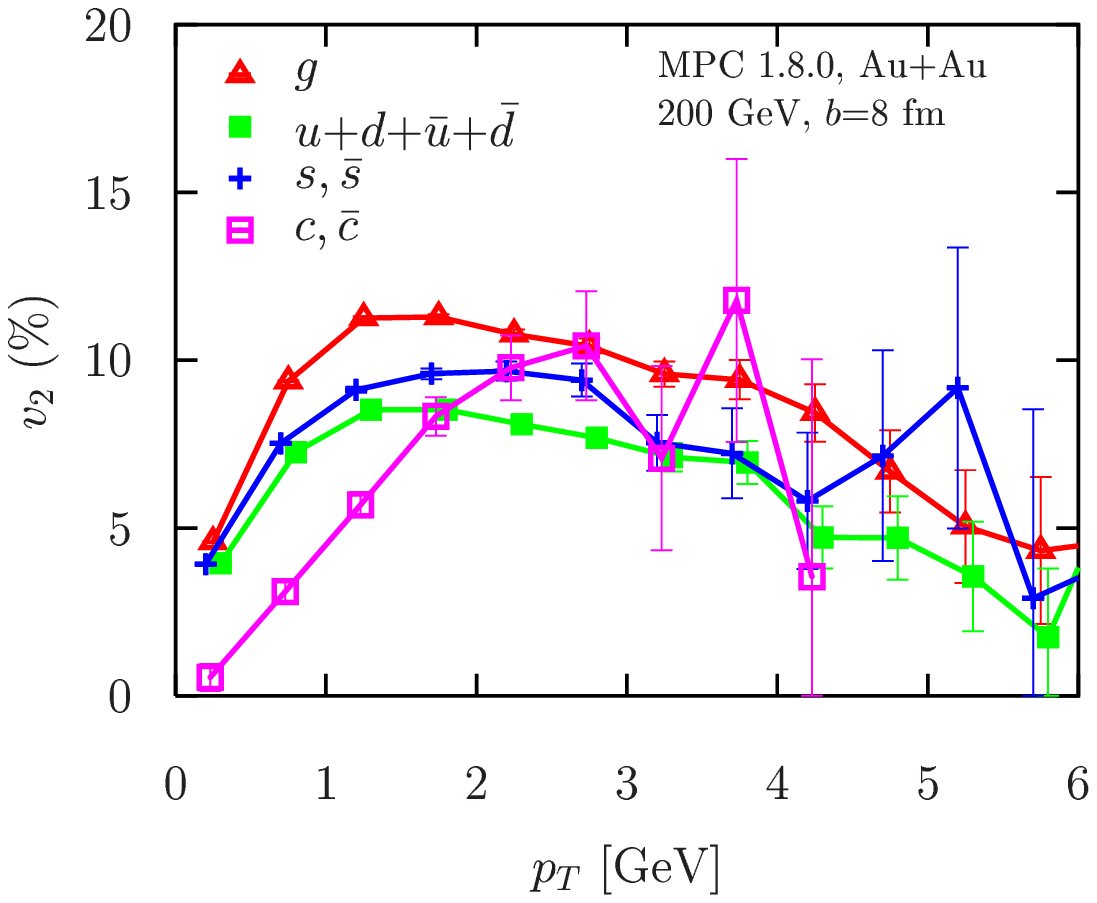,height=2.1in,width=2.6in,clip=5,angle=0} 
\hspace*{-0.2cm}%
\epsfig{file=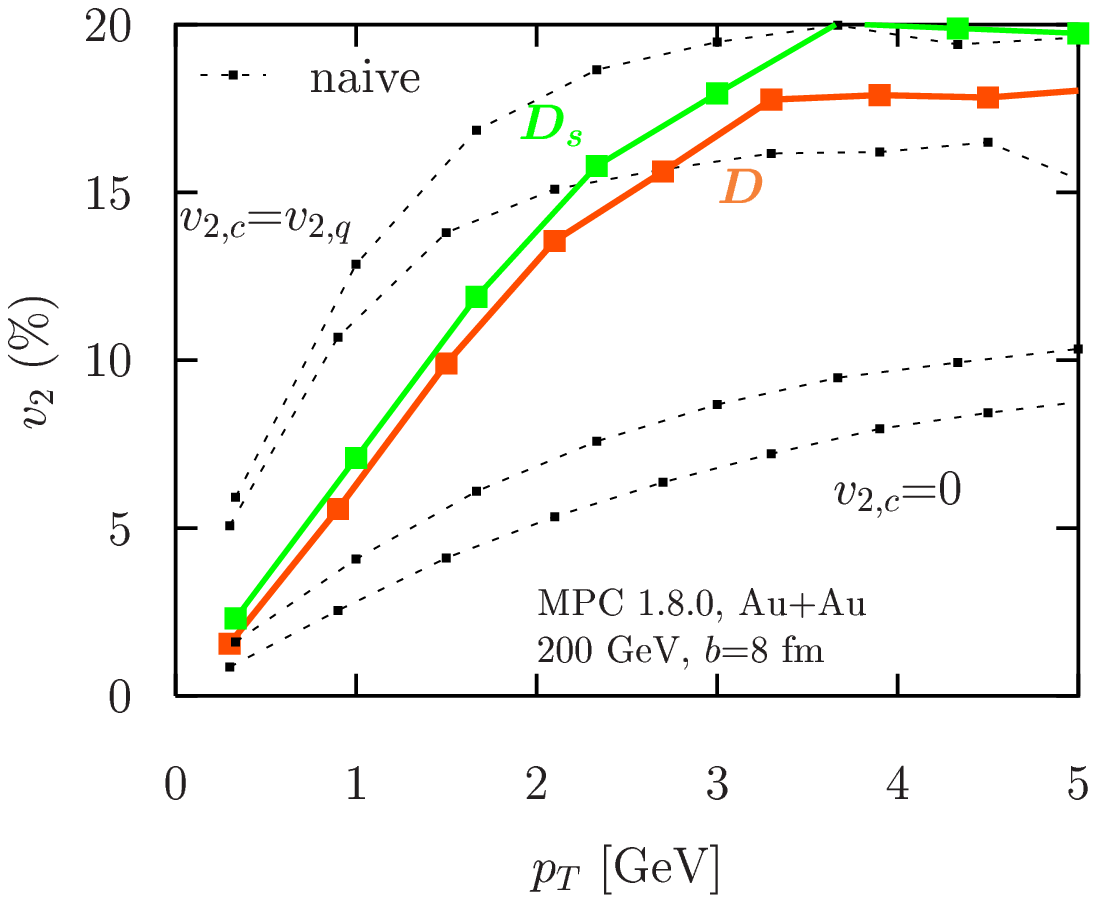,height=2.1in,width=2.6in,clip=5,angle=0} 
\vspace*{-0.2cm} 
\caption{\label{fig2}
Final parton elliptic flows (left panel) and $D$ meson elliptic flow 
(right panel) as a function of $p_T$ 
in $Au+Au$ at $\sqrt{s_{NN}}=200$ GeV with $b=8$ fm at RHIC, 
computed using the transport model MPC~\cite{MPC}. 
The $D$ meson $v_2$ was obtained from the parton flows
via the quark coalescence formula in~\cite{charmcoal}.
The dashed lines correspond to the two extreme scenarios in~\cite{charmcoal}
(see text).
For hadronization via independent fragmentation, 
charm hadron $v_2$ is approximately the same as the charm quark $v_2$.
}
\end{figure}

On the other hand, the elliptic flow evolution is driven mainly by
elastic scatterings. Scatterings off gluons $gX \to gX$ 
are the most relevant because gluons are most abundant and also have
larger cross sections (\ref{el_xsec}).
Figure~\ref{fig2}a shows the elliptic flow of partons at freezeout 
as a function of $p_T$.
For light partons, $v_2(p_T)$ reaches  $\approx 10$\%, 
in quantitative agreement with \cite{v2}.
However, instead of saturation between $p_T\sim 1-6$ GeV, 
this calculation shows a turnover and slow decrease 
of $v_2$ at high $p_T$. The reason for the difference is
that Ref.~\cite{v2} considered thermal initial conditions,
while here the initial spectra are not thermal.
The slight difference in Fig.~\ref{fig2}a 
between strange and light quark $v_2$
is due to the flavor dependence of the initial minijet distributions.
Because of the larger gluon cross sections (\ref{el_xsec}),
the gluon $v_2$ exceeds the quark flows.

At low $p_T \lton 2$ 
GeV charm quark elliptic flow is suppressed 
relative to the $v_2$ of light partons, 
confirming expectations that heavy quark momenta are more difficult to 
randomize.
This behavior is reminiscent of the reduction of elliptic flow for heavy
particles found in 
ideal hydrodynamics~\cite{Kolbhydro,Huovinenhydro,Teaneyhydro,Hiranohydro}.
In contrast, at high $p_T > 3$ GeV, 
charm quark $v_2$ is about the same as the light
quark $v_2$. 
At large momenta, the heavy-light difference should disappear
because the particle mass cannot play a role as $E/M\to \infty$.
Remarkably, the transport solutions show that most of 
the difference disappears already by
$p_T \sim 2.5 - 3 $ GeV, i.e., when $p_T/M_c \sim$ a few.

\section{Charm hadron elliptic flow}

From the elliptic flow of partons, one can predict the hadron
elliptic flows using a suitable hadronization model.
Here we are interested in the elliptic flow of the $D$ and $D_s$,
which will likely be the first charm hadrons measured with sufficient
statistics at high $p_T$ at RHIC.

A simple hadronization model is independent fragmentation, in which
each parton fragments independently to hadrons. 
In the collinear approximation commonly employed,
hadron momenta are related to the parton momentum
via $\vp_{T,h} = z \,\vp_{T,parton}$, 
where $z$ is distributed according to the fragmentation function $D(z)$.
Because for heavy quarks $D(z)$ strongly peaks around 
$z\approx 1$~\cite{Peterson},
{\em charm hadron elliptic flow 
is essentially identical to the charm quark $v_2$} shown in Fig.~\ref{fig2}a.

On the other hand, there are strong
indications~\cite{STARv2identB,PHENIXv2identA,STARv2identC}
that at intermediate $2 \lton p_T \lton 5$ GeV
hadronization may dominantly occur via quark 
coalescence~\cite{Dukecoal,Texascoal,VoloshinQM02,coalv2,Hwacoal}.
In the quark coalescence approach, a constituent quark and an antiquark
that are close in phasespace can combine to form a meson, $\alpha\beta \to M$,
while three constituent (anti)quarks can form an (anti)baryon, 
$\alpha\beta\gamma\to B$.

In the simplest (but most successful)
variant of the model~\cite{VoloshinQM02,coalv2,charmcoal}, 
hadron elliptic flow is approximately
the sum of constituent flows
\be
\fl
v_{2,B}(x,p_T) \approx \!\! \sum\limits_{i=\alpha,\beta,\gamma} \!\!
                   v_{2,i}(x,p_{T,i}) \ , \qquad
v_{2,M}(x,p_T) \approx \!\! \sum\limits_{i=\alpha,\beta} v_{2,i}(x,p_{T,i}) 
\ ,
\label{v2scaling}
\ee
with corrections ${\cal O}(\{v_{2,i}\}^3)$ that are small in our case.
Here $\sum p_{T,i} = p_T$, and the hadron 
momentum is shared roughly in proportion to constituent mass~\cite{charmcoal}.
For hadrons composed of $u$, $d$, and $s$ quarks,
the sharing is approximately equal,
while for $D$ mesons or the $\Lambda_c$,
the heavy quark carries most of the momentum.

Figure~\ref{fig2}b shows $D$ and $D_s$ elliptic flow as a function of $p_T$ 
given by
(\ref{v2scaling}), for constituent mass ratios $m_{u,d} : m_c = 1 : 5$,
and $m_s : m_c = 1 : 3$. The results are compared to two extreme 
scenarios~\cite{charmcoal}: i) zero charm $v_2$, i.e.,
no rescatterings at all (lower dashed lines); and ii) 
$v_2^{charm}(p_T) = v_2^{light}(p_T)$ (upper dashed lines),
which is equivalent to the assumption that $D$ flows the same way as all light
mesons $v_2^{D}(p_T) \approx v_2^{K}(p_T) \approx v_2^{\pi}(p_T)$.
Neither of the two extremes applies in general.
At high $p_T > 2.5-3$ GeV, $v_{2,c} \approx v_{2,q}$ and hence 
the situation agrees with scenario ii),
while at low $p_T$ it is closer to i) because of the suppression
in the charm quark $v_2$.
Thus, 
$D$ meson elliptic flow is predicted to rise smoothly at low $p_T$, 
and to {\em saturate at $\approx 50\%$
higher $p_T$ than the pion and kaon $v_2$
but at the same magnitude}.

The above results ignore fragmentation contributions and therefore
are not valid at very high $p_T$.
For light quarks, the region of validity was estimated~\cite{Dukev2} to be
$p_T \lton 4-5$ GeV. For charm, the window is smaller because:
i) while for light quarks coalescence ``amplifies'' 
transverse momenta by factor two to three,
for charm the hadron $p_T$ is only $20-30$\% larger than 
the charm quark $p_T$;
ii) fragmentation is harder for charm quarks than for lighter partons,
therefore charm hadron spectra from fragmentation
do not fall as steeply at high $p_T$;
and iii) radiative energy loss (jet quenching) is expected to be smaller for 
charm quarks~\cite{heavyqDK,heavyqDG} 
than for light partons, therefore
the fragmentation contribution is less suppressed at high $p_T$.
Thus,
it is not likely that the coalescence results are valid above
$p_T \sim 3-4$ GeV.

Finally we emphasize that,
despite the success of the simple quark coalescence models, 
severe problems arise
when spacetime inhomogeneities or dynamical aspects
are considered~\cite{dyncoal,coalscaling}.
For example, surface-like emission of
high-$p_T$ particles (as seen from covariant transport theory)
results in large, spatially nonuniform local momentum anisotropies,
for which the scaling (\ref{v2scaling}) requires
a high degree of lucky cancellations to occur~\cite{coalscaling}.

Furthermore, in the dynamical coalescence approach of Ref.~\cite{dyncoal},
elliptic flow scaling (\ref{v2scaling}) is violated
for light quarks.
That model combines covariant parton transport theory 
with the coalescence formula~\cite{GFR} by Gyulassy, Frankel and Remler
that is applicable to diffuse 4D freezeout distributions in spacetime.
Constituents without a coalescence partner are assumed to fragment
independently.
Because a significant fraction of constituents has no partner near enough
in phasespace to coalesce with,
elliptic flow is reduced relative to the flow scaling expectation.

\begin{figure}[hbpt] 
\center
\hspace*{-0.2cm}
\epsfig{file=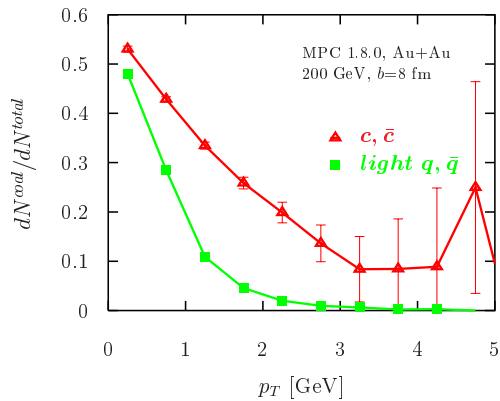,height=2.1in,width=2.6in,clip=5,angle=0}
\vspace*{-0.2cm} 
\caption{\label{fig3}
Fraction of charm quarks that hadronize via coalescence as opposed to
fragmentation, as a function $p_T$ 
in $Au+Au$ at $\sqrt{s_{NN}}=200$ GeV with $b=8$ fm at RHIC.
Computed in a dynamical coalescence approach~\cite{dyncoal}
using the transport model MPC~\cite{MPC} and the Gyulassy-Frankel-Remler
coalescence formula~\cite{GFR}.
}
\end{figure}

The same problem would be present for charm quarks in that approach,
as illustrated in Fig.~\ref{fig3}.
The fraction of light and charm
constituents that hadronize via coalescence as opposed to 
fragmentation are shown as a function of $p_T$.
Above $p_T > 3$ GeV, less than $25$\% of charm quarks 
comes from coalescence
in the dynamical approach, resulting in 
a reduction of $D$ meson elliptic flow by almost a half
above $p_T \approx 3.5$ GeV.

\section{Conclusions}

This study is the first calculation
of charm elliptic flow in $Au+Au$ at $\sqrt{s_{NN}} \sim 200$ GeV
at RHIC from covariant parton transport theory.
Both elastic $2\to 2$ and inelastic $2\to 2$ parton interactions
were included using MPC~\cite{MPC}.
We show that parton opacities
needed to generate a $\approx 10$\% light quark elliptic flow
also imply a charm quark elliptic flow of similar magnitude, at
moderately large $p_T > 2.5-3$ GeV.
At lower transverse momenta charm $v_2$ is suppressed,
analogously to the mass dependence of elliptic flow 
found in ideal hydrodynamics.
Based on hadronization via quark coalescence,
we predict that the elliptic flows of $D$ and $D_s$
saturate at the same magnitude $\approx 20$\% as pion and kaon $v_2$
but at $p_T \approx 3$ GeV, i.e., $50$\% higher 
than the ``meson value'' of $2$ GeV.
In contrast, from hadronization via independent fragmentation,
charm hadron elliptic flow is only $\approx 10$\% (same as charm quark $v_2$).

We emphasize that the above quark coalescence prediction relies on the 
approximate additivity of elliptic flow, 
which agrees well with observations in the light sector but
is quite problematic to preserve in a dynamical coalescence 
approach~\cite{dyncoal,coalscaling}.
In any case, above $p_T \sim 3-4$ GeV, 
fragmentation contributions are expected to dominate
and reduce $D$ and $D_s$ $v_2$ to $\approx 10$\%.

Finally, the large charm elliptic flow is accompanied by significant flavor 
equilibration at RHIC, with an about five-fold enhancement of strangeness 
and a $40-50$\% secondary production of charm.

\ack
Computer resources by the PDSF/LBNL are gratefully acknowledged. 
This work was supported by DOE grant DE-FG02-01ER41190.


\section*{References}
 
\begin{thebibliography}{99}
 

\bibitem{flow-review}
For reviews see, e.g.,
Ollitrault J
1998 {\it  Nucl.\ Phys.\ } A {\bf 638} 195;
Poskanzer A M
2001 {\it Preprint} nucl-ex/0110013;
Hirano T
2004 {\it J.\ Phys.\ } G {\bf 30} S845; or Heinz~U and Kolb~P 2004 
{\it Preprint} nucl-th/0305084



\bibitem{Binv2}
Zhang B, Gyulassy M and Ko C M
1999 {\it Phys.\ Lett.\ }B {\bf 455} 45

\bibitem{v2}
Molnar D and Gyulassy M
2002 {\it Nucl.\ Phys.\ }A {\bf 697} 495;
2002 {\it ibid.} {\bf 703} 893(E); 2002 {\it ibid.} {\bf 698} 379

\bibitem{pQCDv2}
Wang X 
2001 {\it Phys.\ Rev.\ }C {\bf 63} 054902;
Gyulassy M, Vitev I and Wang X N
2001 {\it Phys. Rev. Lett.} {\bf 86} 2537;
Gyulassy M {\it et al.}
2002 {\it Phys.\ Lett.\ }B {\bf 526} 301

\bibitem{coalv2}
Molnar D and Voloshin S A
2003 {\it Phys.\ Rev.\ Lett.\  } {\bf 91} 092301;
Molnar D 2003 {\it J.\ Phys.\ } G {\bf 30} S235


\bibitem{thermalFits}
Braun-Munzinger P {\it et al.}
2001 {\it Phys.\ Lett.\ } B {\bf 518} 41

\bibitem{Kolbhydro}
Kolb P F {\it et al.}
2001 {\it Nucl.\ Phys.\ } A {\bf 696} 197;
2001 {\it Phys.\ Lett.\ } B {\bf 500} 232

\bibitem{Teaneyhydro}
D.~Teaney, J.~Lauret and E.~V.~Shuryak,
nucl-th/0110037.

\bibitem{Hiranohydro}
Hirano~T and Tsuda~K 2002
{\it Phys.\ Rev.\ }C {\bf 66} 054905


\bibitem{PHENIXe}
Adcox K {\it et al.}  [PHENIX Collaboration],
2002 {\it Phys.\ Rev.\ Lett.\  } {\bf 88} 192303

\bibitem{Batsouli}
Batsouli S {\it et al.}
2003 {\it Phys.\ Lett.\ } B {\bf 557} 26


\bibitem{Kaneta}
Kaneta~M  [PHENIX Collaboration] 2004
{\it J.\ Phys.\ }G {\bf 30} S1217

\bibitem{charmcoal}
Lin~Z~w and Molnar~D 2003
{\it Phys.\ Rev.\ }C {\bf 68} 044901

\bibitem{Texascharm}
Greco~V, Ko~C~M, Rapp R
2004 {\it Phys.\ Lett.\ } B {\bf 595} 202

\bibitem{ZPC}
Zhang B
1998 {\it Comput.\ Phys.\ Commun.\  }{\bf 109} 193

\bibitem{nonequil}
Molnar D and Gyulassy M
2000 {\it Phys.\ Rev.\ }C {\bf 62} 054907

\bibitem{inelv2}
%
Molnar D 
1999 {\it Nucl.\ Phys.} {\bf A661} 236

\bibitem{hbt}
Molnar D and Gyulassy M
2004 {\it Phys.\ Rev.\ Lett.\ } {\bf 92} 052301

\bibitem{dyncoal}
Molnar D {\it Preprint} 2004 nucl-th/0406066;
2004 {\it J.\ Phys.\ } G {\bf 30}, S1239

\bibitem{XuGreiner}
Xu~Z and Greiner~C 2004
{\it Preprint} hep-ph/0406278

\bibitem{ALCORMICOR}
Biro T S, Levai P and Zimanyi J
1995 {\it Phys.\ Lett.\ }B {\bf 347} 6;
Csizmadia P and Levai P
2002 {\it J.\ Phys.\ }G {\bf 28} 1997


\bibitem{VoloshinQM02}
Voloshin~S~A 2003
{\it Nucl.\ Phys.\ }A {\bf 715} 379


\bibitem{Texascoal}
Greco~V, Ko~C~M and Levai~P 2003
{\it Phys.\ Rev.\ Lett.\ } {\bf 90} 202302;
2003 {\it Phys.\ Rev.\ }C {\bf 68} 034904

\bibitem{Dukecoal}
Fries~R~J {\it et al.} 2003
{\it Phys.\ Rev.\ Lett.\ } {\bf 90} 202303;
Fries~R~J {\it et al.} 2003
{\it Phys.\ Rev.\ }C {\bf 68} 044902

\bibitem{Hwacoal}
Hwa R C and Yang C B
2002 {\it Phys.\ Rev.\ } C {\bf 66} 025205;
2003 {\it ibid.} {\bf 67} 034902

\bibitem{coalscaling}
Molnar D 2004 {\it Preprint} nucl-th/0408044


\bibitem{Kolbcoal}
Kolb~P~F {\it et al.} 2004
{\it Phys.\ Rev.\ }C {\bf 69} 051901

\bibitem{Combridge}
Combridge B L 1979 {\it Nucl.\ Phys.\ } {\bf B151} 429

\bibitem{MPC}
Molnar D 2004 {\it Computer code} MPC~1.8.0.
This parton cascade code can be downloaded from  WWW at
http://www-cunuke.phys.columbia.edu/people/molnard

\bibitem{subdivision}
Pang Y 1996 {\it Preprint} RHIC 96 Summer Study, CU-TP-815 (unpublished)

\bibitem{Huovinenhydro}
Huovinen~P {\it et al.} 2001
{\it Phys.\ Lett.\ }B {\bf 503} 58

\bibitem{Peterson}
Peterson C {\it et al.}
1983 {\it Phys.\ Rev.\ } D {\bf 27} 105

\bibitem{STARv2identB} 
Adams~J {\it et al.}  [STAR Collaboration] 2004
{\it Phys.\ Rev.\ Lett.\ } {\bf 92} 052302;
Adler~C {\it et al.}  [STAR Collaboration] 2002
{\it Phys.\ Rev.\ Lett.\ } {\bf 89} 132301

\bibitem{PHENIXv2identA} 
Adler~S~S {\it et al.}  [PHENIX Collaboration] 2003
{\it Phys.\ Rev.\ Lett.\ } {\bf 91} 182301

\bibitem{STARv2identC} 
Castillo J [STAR Collaboration]
2004 {\it J.\ Phys.\ } G {\bf 30} S1207

\bibitem{Dukev2}
Nonaka C {\it et al.}
2004 {\it Phys.\ Lett.\ } B {\bf 583} 73

\bibitem{heavyqDK}
Dokshitzer Y L and Kharzeev D E
2001 {\it Phys.\ Lett.\ } B {\bf 519} 199

\bibitem{heavyqDG}
Djordjevic M and Gyulassy M
2003 {\it Phys.\ Lett.\ }B {\bf 560} 37;
2004
{\it J.\ Phys.\ }G {\bf 30}, S1183


\bibitem{GFR}
Gyulassy M, Frankel K  and Remler E a 
1983 {\it Nucl.\ Phys.\ }A {\bf 402} 596

\end{thebibliography}
\end{document}